\documentclass[useAMS]{mn2e}

\usepackage{txfonts}
\usepackage{graphicx}
\usepackage{url}

\newcommand\ion[2]{{\sc #1}\,{\sc #2}}

\title[Linear polarization in 3C\,84]{A search for linear polarization in the active galactic nucleus 3C\,84 at 239 and 348\,GHz\thanks{This study is based on observations carried out with the IRAM Plateau de Bure Interferometer. IRAM is supported by INSU/CNRS (France), MPG (Germany), and IGN (Spain).}}
\author[S. Trippe et al.]{
S. Trippe$^{1}$\thanks{E-mail: {\tt trippe@astro.snu.ac.kr}},
M. Bremer$^{2}$,
T. P. Krichbaum$^{3}$,
M. Krips$^{2}$,
R. Neri$^{2}$,
V. Pi\'etu$^{2}$,\newauthor\
J. M. Winters$^{2}$
\\
$^{1}$Seoul National University, Department of Physics and Astronomy, 599 Gwanak-ro, Gwanak-gu, Seoul 151-742, South Korea\\
$^{2}$Institut de Radioastronomie Millim\'etrique (IRAM), 300 rue de la Piscine, F-38406 Saint Martin d'H\`eres, France\\
$^{3}$Max-Planck-Institut f\"ur Radioastronomie, Auf dem H\"ugel 69, D-53121 Bonn, Germany
}
\begin{document}

\date{Accepted 2012 June 9. Received 2012 June 8; in original form 2012 March 30}

\pagerange{\pageref{firstpage}--\pageref{lastpage}} \pubyear{2012}

\maketitle

\label{firstpage}

\begin{abstract}
We report a search for linear polarization in the active galactic nucleus (AGN) 3C\,84 (NGC~1275) at observed frequencies of 239\,GHz and 348\,GHz, corresponding to rest-frame frequencies of 243\,GHz and 354\,GHz. We collected polarization data with the IRAM Plateau de Bure Interferometer via Earth rotation polarimetry. We do not detect linear polarization. Our analysis finds $3\sigma$ upper limits on the degree of polarization of 0.5\% and 1.9\% at 239\,GHz and 348\,GHz, respectively. We regard the influence of Faraday conversion as marginal, leading to expected circular polarizations $\lesssim$0.3\%. Assuming depolarization by a local Faraday screen, we constrain the rotation measure, as well as the fluctuations therein, to be $\gtrsim10^6$\,rad\,m$^{-2}$. From this we estimate line-of-sight magnetic field strengths of $\gtrsim100\mu$G. Given the physical dimensions of 3C\,84 and its observed structure, the Faraday screen appears to show prominent small-scale structure, with $\Delta{\rm RM}\gtrsim10^6$\,rad\,m$^{-2}$ on projected spatial scales $\lesssim$1\,pc.

\end{abstract}


\section{Introduction}

At radio frequencies, active galactic nuclei (AGN) are luminous emitters of synchrotron radiation (see, e.g., Krolik \cite{krolik1999}, or Kembhavi \& Narlikar \cite{kembhavi1999}, for reviews). Radio observations of AGN find degrees of linear polarizations in the range $\approx$1--20\% with a mean value of about $\approx$5\% (Altschuler \& Wardle \cite{altschuler1976,altschuler1977}; Aller et al. \cite{aller1985}; Nartallo et al. \cite{nartallo1998}; Trippe et al. \cite{trippe2010}; Agudo et al. \cite{agudo2010}). Linear polarization provides information on physical parameters of emitters like emission region structure, geometry and strengths of magnetic fields, and particle densities (e.g., Saikia \& Salter \cite{saikia1988}, and references therein). Accordingly, studies of their polarized light deepen the understanding of the physics of active galaxies.

The active nucleus of the Seyfert~2 galaxy 3C\,84 (NGC~1275), located at a redshift $z$=0.018, is known for its highly unusual radio polarization properties. Linear polarization has been found on levels consistently lower than 1\% at frequencies from 5\,GHz to 43\,GHz\footnote{According to, e.g., monitoring results from the Effelsberg radio observatory (U. Bach, MPIfR Bonn, priv. comm.) or the Very Large Array calibration data base (\url{http://www.aoc.nrao.edu/~smyers/calibration/master.shtml}).}. Radio-interferometric maps obtained with the Very Long Baseline Array (VLBA) at 15\,GHz, providing angular resolutions of $\approx$2 milliseconds of arc, unveil linear polarization on levels $\lesssim$8\% within a small region less than about one parsec in size (Taylor et al. \cite{taylor2006})\footnote{See also, e.g., the data bases of the MOJAVE survey (Lister et al. \cite{lister2009}; \url{http://www.physics.purdue.edu/astro/MOJAVE/sourcepages/0316+413.shtml}) or the University of Michigan Radio Astronomy Observatory (\url{http://www.astro.lsa.umich.edu/obs/radiotel/umrao.php}).}. Additionally, 3C\,84 shows substantial \emph{circular} polarization with degrees of circular polarization being $\approx$0.2--3\% depending on frequency and angular resolution (Aller et al. \cite{aller2003}; Homan \& Wardle \cite{homan2004}; Agudo et al. \cite{agudo2010}). The combination of low linear and high circular polarization indicates efficient Faraday depolarization (e.g., Dreher et al. \cite{dreher1987}) and Faraday conversion (e.g., Jones \cite{jones1988}) by matter surrounding the nucleus. Accordingly, Taylor et al. \cite{taylor2006} identify ionized gas permeated by magnetic fields with substructure on scales $\lesssim$10\,pc as Faraday screen.

Given the state of the art, a detection and analysis of linear polarization in 3C\,84 may be achieved via two roads. First, high angular resolution mapping is apparently able to separate high and low polarization regions spatially and reduce the ``masking'' of polarized flux by spatial averaging in telescope beams. Second, observations at high (sub-)millimetre radio frequencies should be less affected by Faraday depolarization than previous observations aimed at lower frequencies. This is a consequence of the fact that the strength of Faraday depolarization, and thus the loss of linear polarization, is proportional to the square of the rest-frame wavelength $\lambda_0^2$.

In this article we report the results of observations with the IRAM Plateau de Bure Interferometer (PdBI; Winters \& Neri \cite{winters2011}\footnote{\url{http://www.iram.fr/IRAMFR/GILDAS/doc/pdf/pdbi-intro.pdf}}) at 239\,GHz and 348\,GHz. We followed the high-frequency approach outlined above in order to detect the linear polarization of 3C\,84 without resolving the source spatially.

\section{Observations and Data Analysis}

The Plateau de Bure Interferometer is composed of six antennas with 15\,m diameter each. All antennas are equipped with dual linear polarization Cassegrain focus receivers. Both orthogonal linear polarizations -- ``horizontal'' (H) and ``vertical'' (V) with respect to the antenna frame -- are observed simultaneously. Observations can be carried out (non-simultaneously) in four atmospheric windows located around wavelengths of 0.8\,mm, 1.3\,mm, 2\,mm, and 3\,mm.   Each of these bands covers a continuous range of frequencies. Frequency ranges are
277--371\,GHz for the 0.8-mm band,
201--267\,GHz for the 1.3-mm band,
129--174\,GHz for the 2-mm band,
and 80--116\,GHz for the 3-mm band.
Within a given band, any frequency is available for observations.

At the time of the observations presented here, the PdBI was not yet equipped for observations of all Stokes parameters. We obtained linear polarization data via Earth rotation polarimetry, i.e. by monitoring the fluxes in the H and V channels as functions of parallactic angle $\psi$. For deriving the polarization of a source, we calculate the parameter

\begin{equation}
q(\psi) = \frac{V-H}{V+H}(\psi) = \frac{Q}{I}\cos(2\psi) + \frac{U}{I}\sin(2\psi)
\label{eq_q}
\end{equation}

\noindent
from the fluxes $H(\psi)$ and $V(\psi)$. Here $I$, $Q$, and $U$ are the relevant Stokes parameters. The second equality means that $q(\psi)$ provides full information on linear polarization (see, e.g., Sault, Hamaker \& Bregman \cite{sault1996}; Thompson, Moran \& Swenson \cite{thompson2001}; but also Beltr\'an et al. \cite{beltran2004}) if a sufficient range of $\psi$ is observed.

Due to the nature of polarized light and the fact that the PdBI antenna receivers are located in the Cassegrain foci, observing a polarized target results in $q(\psi)$ being a cosinusoidal signal with a period of $180^{\circ}$. The functional form of $q(\psi)$ is thus

\begin{equation}
q(\psi) \equiv m_L\cos\left[2(\psi-\chi)\right] ~ .
\label{eq_polformula}
\end{equation}

\noindent
Here $m_L$ is the fraction of linear polarization (ranging from 0 to 1; in the following, we will express $m_L$ in units of \%) and $\chi$ is the polarization angle (ranging from 0$^{\circ}$ to 180$^{\circ}$). A more detailed discussion of the methodology is provided in Trippe et al. \cite{trippe2010}.

\begin{figure}
\includegraphics[angle=-90,width=83mm]{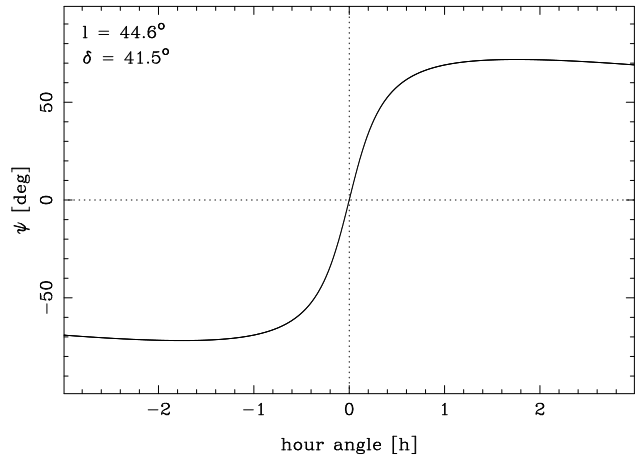}
\caption{Parallactic angle $\psi$ (in units of degrees) vs. hour angle (in units of hours) for 3C\,84 as observed from the latitude of the PdBI. The source declination is $\delta=41.5^{\circ}$, the latitude of the observatory is $l=44.6^{\circ}$.}
\label{fig_psi}
\end{figure}

As stated by Eq.~\ref{eq_q}, $q(\psi)$ is a relative parameter and therefore not affected by inaccuracies of flux or gain calibrations -- equal factors multiplied to H and V cancel out. Accordingly, fluctuations of gain or atmospheric opacity do not affect $q(\psi)$ systematically. Equally, we do not require a dedicated calibration of amplitude or flux scales. However, we identify two sources of systematic errors we need to address before proceeding with the analysis (see also the corresponding discussion in Trippe et al. \cite{trippe2010}).

First, we may expect a certain level of instrumental polarization. As for the PdBI, the receivers are located in the Cassegrain foci of the antennas. Accordingly, the instrumental polarization signal is fix with respect to the antenna, whereas the astronomical polarization signal is fix with respect to the sky. In terms of $q(\psi)$, astronomical polarization introduces a cosinusoidal dependence on $\psi$ as stated by Eq.~\ref{eq_polformula}. Instrumental polarization contributes a constant offset $o$, meaning we effectively deal with a modified polarization model

\begin{equation}
q(\psi) \longrightarrow q'(\psi) = m_L\cos\left[2(\psi-\chi)\right] + o ~ .
\label{eq_pol}
\end{equation}

\noindent
The exact value of instrumental polarization is not well known for the PdBI. Trippe et al. \cite{trippe2010} were able to probe the behaviour of the PdBI receivers in the frequency range 84--116\,GHz by means of laboratory experiments. For this frequency range, they found receiver polarizations of $\sim$0.3\%. However, astronomical and instrumental polarization appear as separate parameters in Eq.~\ref{eq_pol}. This means that instrumental polarization is not able to modify the values we observe for $m_L$ in a systematic fashion. However, this statement does not hold entirely if instrumental polarization is a function of antenna orientation; we discuss this effect in detail in Sect.~3.

Second, there may be systematic differences in the gains, or efficiencies, of the H and V channels. If this is the case, one can rewrite the H channel flux as $H(\psi)\rightarrow rH(\psi)$ with $r>0$ being the ratio of the gains of channels H and V. Any $r\neq1$ modifies $q(\psi)$ such that (1) the mean value of $q(\psi)$ differs from zero, and (2) the polarization amplitude $m_L$ is slightly reduced. Effect (1) is absorbed by the parameter $o$ in Eq.~\ref{eq_pol}. Effect (2) is very small even for large $r=0.8...1.2$. As one can easily evaluate numerically, such gain ratios lead to a relative reduction of degrees of polarization $m_L$ by $\approx$1\% at most. This means that for a source with a physical $m_L=1$\% we would actually measure $m_L'=0.99$\% -- an effect we may safely ignore. Eventually, we conclude that both, instrumental polarization as well as gain differences, are kept under control by use of polarization model $q'(\psi)$ (Eq.~\ref{eq_pol}).

\begin{figure*}
\includegraphics[width=175mm]{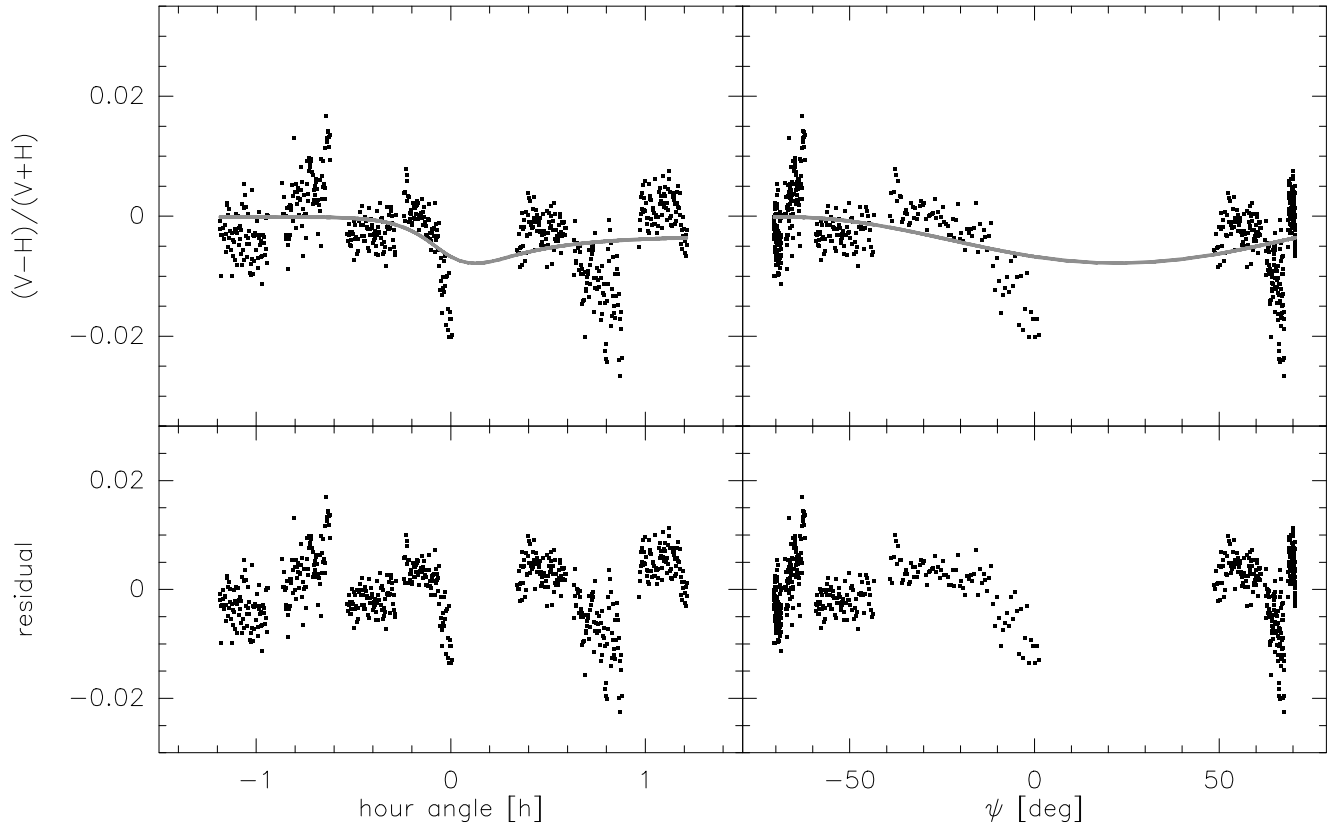}
\caption{Polarization analysis for 239\,GHz flux data. The top panels show $q=(V-H)/(V+H)$ as function of hour angle (on the left) and parallactic angle $\psi$ (on the right). Points indicate data, grey curves indicate the best-fitting polarization models according to Eq.~\ref{eq_polformula}. The bottom panels show the corresponding residuals, meaning the differences between data and best-fitting model curves.}
\label{fig_pol239}
\end{figure*}

We observed 3C\,84 on 9 March 2011 at a frequency of 348\,GHz and on 11 March 2011 at 239\,GHz. We recorded source fluxes in cycles of 15 minutes composed of 30 adjacent ``scans'' of 30 seconds duration each. Each scan is divided further into 30 ``dumps'' of one second duration each. Between two consecutive cycles, pointing and focus of the antennas were checked and adjusted if necessary, using 3C\,84 as reference point source; these calibration intervals took between three and seven minutes of time. Accurate Earth rotation polarimetry requires a good coverage in parallactic angle. Therefore we executed all observations in time windows centred on the transit. The PdBI is located at a latitude $l=44.6^{\circ}$ and 3C\,84 has a declination $\delta=41.5^{\circ}$. The parallactic angle $\psi$ as function of hour angle follows the pattern presented in Fig.~\ref{fig_psi}. The extreme values $\psi=\pm72^{\circ}$ are located at hour angles $\pm1.8$\,h. The $\psi$ curves varies only very slowly at times more than about one hour before or after the transit. As this leads to a clustering of $q(\psi)$ values around $\psi\approx70^{\circ}$ which can distort a fit of our model (Eq.~\ref{eq_polformula}) to the data, we actually limit our analysis to hour angles $<$1.25\,h from the transit, meaning $|\psi|\leq71^{\circ}$.

In both observing runs (9 and 11 March 2011) we observed 3C\,84 with five of the six antennas of the PdBI. We obtained antenna-based amplitudes and phases from factorization of the complex visibilities (see, e.g., Sect.~7.5 of Fomalont \& Perley \cite{fomalont1999}). Gaps due to calibration intervals and technical interruptions aside, we stored flux data\footnote{We processed the amplitudes separately, in order not to be affected by atmospheric phase fluctuations. Ignoring the phase information is possible because 3C\,84 is a point source for the PdBI.} from each antenna with time resolutions of one second (corresponding to one dump) for both linear polarizations. We used the PdBI wide-band correlator WideX -- which provides a bandwidth of 3.6\,GHz -- in ``continuum mode'', meaning each dump value corresponds to the sum of the flux over the full 3.6\,GHz spectral band. In order to minimize the impact of antenna pointing instabilities, we computed for each polarization a combined lightcurve by taking the median over all antennas, leaving us with two lightcurves for the V and H channels. Thereafter, we binned the flux data in time using a bin size of 10 seconds; this value proved to be a good compromise between dense sampling and low scatter within the lightcurves. Eventually, this procedure left us with 630 and 708 data points per polarization for the 239\,GHz and 348\,GHz lightcurves (for hour angles within $\pm$1.25\,h around the transit), respectively. From the flux data we computed for each observing run $q(\psi)$ according to Eq.~\ref{eq_q}. To derive degrees and angles of polarization, we fit the model $q'(\psi)$ to the data by means of a $\chi^2$ minimization algorithm.\footnote{Throughout this article, $\chi$ always denotes the polarization angle, whereas $\chi^2$ always denotes the weighted sum of the squares of the differences between data and model. This is an unfortunate collision of common nomenclature standards.}

\section{Results}

\begin{figure*}
\includegraphics[width=175mm]{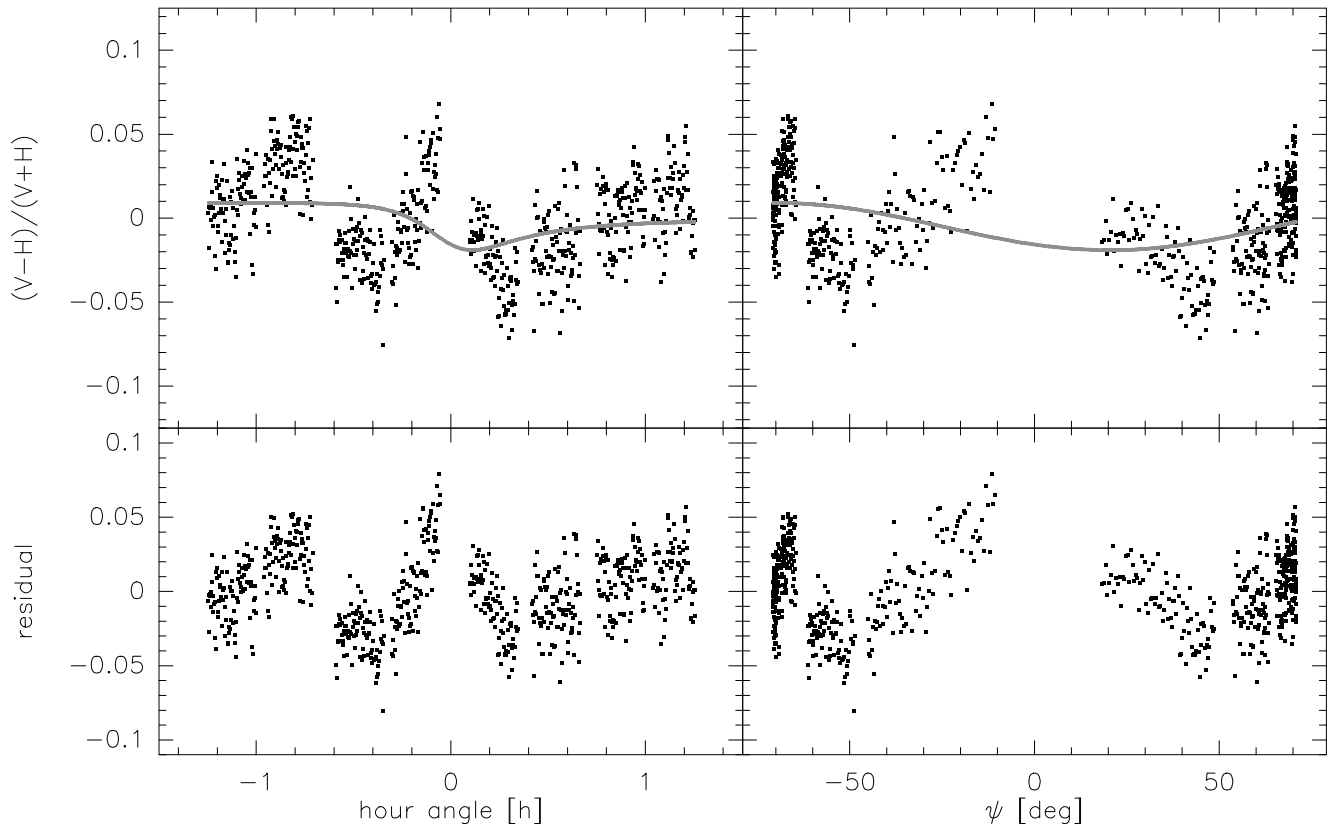}
\caption{Like Fig.~\ref{fig_pol239} albeit for 348\,GHz flux data.}
\label{fig_pol348}
\end{figure*}

We present the results of our analysis in Fig.~\ref{fig_pol239} for the 239\,GHz data and in Fig.~\ref{fig_pol348} for the 348\,GHz data. In both figures we display $q$ as function of parallactic and hour angles, the best-fitting polarization models, and the residuals left when subtracting the models from the data. An obvious feature of the $q$ curves are occasional gaps of various sizes. Our observing scheme requires short calibration intervals every $\approx$15 minutes. Interruptions occurring at or closely around the transit cause substantial gaps in $\psi$ coverage even if they are short in terms of observing time (compare Fig.~\ref{fig_psi}); this is evident in both datasets.

At 239\,GHz, we measured antenna temperatures $T_{\rm A}\approx159$\,mK, translating into a flux density $S_{\nu}\approx6.6$\,Jy for 3C\,84 when adopting a conversion factor of $\rho=41\,\rm Jy\,K^{-1}$. For 1 second of integration time, the statistical measurement uncertainty of the antenna temperature was $\delta T_{A}=2.9$\,mK, corresponding to a relative uncertainty $\delta T_{A}/T_{A}\approx2$\%. At 348\,GHz, we observed $T_{\rm A}\approx71$\,mK, translating into $S_{\nu}\approx4.1$\,Jy when adopting a conversion factor of $\rho=57\,\rm Jy\,K^{-1}$. The statistical measurement uncertainty was $\delta T_{A}=7.2$\,mK, corresponding to a relative uncertainty $\delta T_{A}/T_{A}\approx10$\%.

The accuracy of our observations is limited by systematic uncertainties. In both datasets (Figs. \ref{fig_pol239} and \ref{fig_pol348}) occasional jumps and drifts show up that have amplitudes substantially larger than the random scatter caused by thermal noise. The principal reason for these systematic deviations are pointing instabilities of the PdBI antennas. The PdBI antennas have pointing accuracies of $\approx$2--5'' depending on antenna orientation, quality of the observation of the required reference point source, and weather conditions. An additional complication was provided by the declination of 3C\,84 which is 41.5$^{\circ}$. The PdBI is located at a latitude of 44.6$^{\circ}$, meaning a transit elevation angle of 86.9$^{\circ}$. As the mechanical elevation limit for the PdBI antennas is 87$^{\circ}$, we operated the observatory at the very edge of its mechanical tolerances. The beam width (full width at half maximum) of a PdBI antenna is $\approx$17'' at 239\,GHz and $\approx$12'' at 348\,GHz. Accordingly, pointing deviations by few seconds of arc cause substantial fluctuations in the recorded fluxes. In addition, small pointing differences between the receivers for V and H polarizations artificially create a Stokes $Q\equiv V-H\neq0$. For Gaussian beam profiles, a misalignment of the V and H pointing angles by 10\% (20\%) of the full width at half maximum of the telescope beam corresponds to $|Q|/I\approx2$\%  ($|Q|/I\approx5$\%). These numbers are maximum values assuming that either V or H is pointed ``perfectly'' on the target; the actually observed signal depends on telescope orientation and pointing stability. Indeed our observed $q$ curves show excursions of up to approximately $\pm2$\% and $\pm$5\% at 239\,GHz and 348\,GHz, respectively. This indicates a V-vs-H alignment uncertainty of $\approx$2''.

We fit our polarization model $q'(\psi)$ to the data by means of a $\chi^2$ minimization algorithm. We find degrees of polarization $m_L=0.39\pm0.04$\% and $m_L=1.4\pm0.2$\% for the 239\,GHz and 348\,GHz data, respectively. Errors are statistical. Formally, this corresponds to detections of polarizations on levels $\approx$7--9$\sigma$ which would be highly significant. However, inspection of the residuals indicates a poor agreement of model and data. Therefore we address the significance of the presence (or absence) of a polarization signal by means of an F-test (e.g., M\"uller \cite{muller1975}). An F-test can be used to compare two models applied to the same dataset and to decide if one of them is significantly better than the other. The two models we explore are (1) our polarization model Eq.~\ref{eq_pol} (model ``P'') and (2) a ``null'' model that assumes that the data are intrinsically constant (model ``0''). For both models we calculate the $\chi'^2\equiv\chi^2/{\rm dof}$ (reduced $\chi^2$) of the best fits and from this the parameter

\begin{equation}
f = \frac{\chi'^2_P}{\chi'^2_0}
\label{eq_F}
\end{equation}

\noindent
where $f$ follows an F distribution. The difference between two models is statistically significant on a level $s$ if $f=F_{m,n,s}$. Here $m$, $n$ are the degrees of freedom for the models ``P'' and ``0'', respectively; $s$ is the significance where $s=1-p$ for a false alarm probability $p$. From our analysis we find
\\

$f \equiv F_{627,629,s} = 1.1965$ at 239\,GHz,

$f \equiv F_{705,707,s} = 1.1362$ at 348\,GHz.
\\

\noindent
These values translate into significance levels -- given in fractions as well as Gaussian $\sigma$ -- of
\\

$s = 98.8\% \equiv 2.5\sigma$ at 239\,GHz,

$s = 95.5\% \equiv 2.0\sigma$ at 348\,GHz.
\\

\noindent
Evidently, false alarm probabilities larger than 1\%, corresponding to significance levels $\lesssim2.5\sigma$ in Gaussian terms, indicate that the difference between models ``P'' and ``0'' is insignificant. We therefore conclude that we have not (yet) detected linear polarization signals in our data. Accordingly, we quantify our results by quoting $3\sigma$ upper limits on the degrees of polarization. These limits are given by the $m_L$ values formally derived from the model fits plus three times their statistical errors. The resulting upper limits on $m_L$ are
\\

$m_L < 0.5\%$ at 239\,GHz,

$m_L < 1.9\%$ at 348\,GHz.
\\

\noindent
Obviously, the model fit results for the polarization angles $\chi$ are irrelevant.

\section{Discussion}

Our observations aimed at the analysis of linear polarization in 3C\,84 at observed frequencies of 239\,GHz and 348\,GHz, corresponding to rest-frame frequencies $\nu_0$ of 243\,GHz and 354\,GHz, respectively. Our study complements the observations by Trippe et al. \cite{trippe2010} that found $m_L<1.5\%$ at an observed frequency of 227\,GHz. Accordingly, our study expands the frequency range probed by a factor 1.5, corresponding to a factor 2.3 in $\lambda^2$.

Our observations do not find any indication for linear polarization up to rest-frame frequencies of 354\,GHz. In theory, synchrotron emission is highly polarized: for the case of 3C\,84 with a spectral mm/radio slope -- defined via $S_{\nu}\propto\nu^{-\alpha}$, with $S_{\nu}$ being the flux density -- of $\alpha\approx0.5$ (Trippe et al. \cite{trippe2011}), one may expect degrees of linear polarization of $\approx$69\% and $\approx$12\% for emission from optically thin and optically thick regions, respectively (e.g., Ginzburg \& Syrovatskii \cite{ginzburg1965}; Pacholczyk \cite{pacholczyk1970}). However, these values assume synchrotron emission from homogeneous and isotropic ensembles of electrons moving in uniform magnetic fields. This is an idealization that does not describe properly the situation in AGN that have complex magnetic field structures. In addition, observations suffer from resolution effects: single-dish observations average polarized emission from different source components, leading to ``beam depolarization''. Accordingly, the typical linear polarization from AGN found at radio frequencies is about 5\% (Altschuler \& Wardle \cite{altschuler1976,altschuler1977}; Aller et al. \cite{aller1985}; Nartallo et al. \cite{nartallo1998}; Trippe et al. \cite{trippe2010}; Agudo et al. \cite{agudo2010}).

The very low level of linear polarization in 3C\,84 appears to be connected to the parsec-scale environment of 3C\,84 as well as its geometry. Radio-interferometric maps (VLBA, VLBI) show three main source components: a luminous core, a jet extending about 5\,pc (in projection) to the south, and a counter-jet extending about 3\,pc (in projection) to the north (e.g., Walker \& Anantharamaiah \cite{walker2003}, especially their Fig.~1). The southern jet is directed toward the observer, the northern counter-jet is directed away. The entire galactic nucleus is embedded into dense ionized gas that can act as a Faraday screen (e.g., Heckman et al. \cite{heckman1989}). In addition, the core and the northern jet are located within or behind ionized gas associated with an accretion disk (Walker et al. \cite{walker2000}).

Our observations provide new information on the physical properties of the Faraday screen located in front of the source. Linearly polarized radiation passing through a magnetized plasma experiences Faraday rotation. This means the intrinsic polarization angle $\chi$ is modified like $\chi\longrightarrow\chi'=\chi+\Delta\chi$, with

\begin{equation}
\Delta\chi = {\rm RM}\times\lambda_0^2 ~~ .
\label{eq_pa}
\end{equation}

\noindent
Here $\lambda_0$ is the rest-frame wavelength and RM is the rotation measure

\begin{equation}
{\rm \frac{RM}{rad\,m^{-2}}} = 8.1\times10^5 \int_{\rm l.o.s.} \left(\frac{B_{||}}{\rm G}\right) \left(\frac{n_e}{\rm cm^{-3}}\right) {\rm d}\left(\frac{l}{\rm pc}\right) ~~ ,
\label{eq_rm}
\end{equation}

\noindent
with $B_{||}$ being the strength of the magnetic field parallel to the line of sight (l.o.s.), $n_e$ being the electron number density, and $l$ being the coordinate directed along the l.o.s. (e.g., Rybicki \& Lightman \cite{rybicki1979}; Wilson, Rohlfs \& H\"uttemeister \cite{wilson2010}). If RM shows modulations with amplitudes $\Delta$RM on spatial scales smaller than the source, the source radiation experiences different Faraday rotation depending on the position in the plane of the sky. Observations that do not resolve the RM structure spatially superimpose electric field vectors with different orientations. This partially averages out the polarization signal, reducing the degree of polarization observed. 

Quantitatively, we may understand the strong depolarization of 3C\,84 by describing the Faraday screen as ``Faraday thick'' (e.g., Brentjens \& de Bruyn \cite{brentjens2005}). From Eq.~\ref{eq_pa} it is evident that a medium is Faraday thick if

\begin{equation}
\Delta{\rm RM}\times\lambda_0^2 \gg 1 ~~ .
\label{eq_depth}
\end{equation}

\noindent
Our observations are obtained at wavelengths as short as $\lambda_0=0.85$\,mm. Accordingly, we require $\Delta{\rm RM}\gg10^6$\,rad\,m$^{-2}$ for efficient depolarization.

\begin{figure}
\includegraphics[angle=-90,width=83mm]{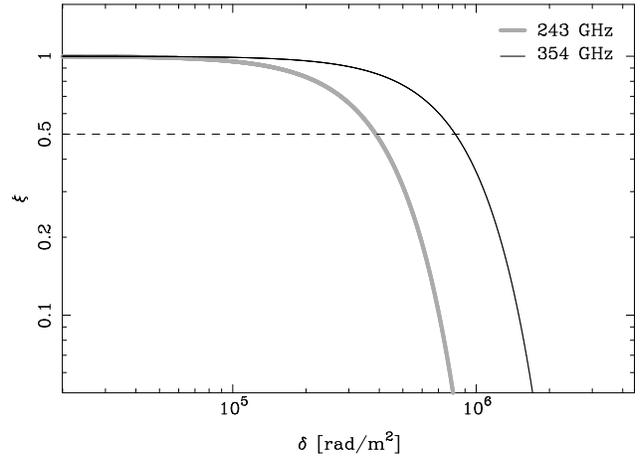}
\caption{Depolarization parameter $\xi$ as function of RM dispersion $\delta$ according to Eq.~\ref{eq_depol}. The grey and black curves indicate the relations for rest-frame frequencies of 243\,GHz and 354\,GHz, respectively. The horizontal dashed line marks the regime of ``substantial'' depolarization, here taken to be $\xi\leq0.5$. } 
\label{fig_depol}
\end{figure}

A more sophisticated calculation is possible when assuming that the RM fluctuations follow a Gaussian distribution with dispersion $\delta\approx\Delta$RM. If the source is not resolved spatially, one finds the depolarization law

\begin{equation}
\xi = \exp\left(-2\delta^2\lambda_0^4\right)
\label{eq_depol}
\end{equation}

\noindent
(Burn \cite{burn1966}; Tribble \cite{tribble1991}). The parameter $\xi\in[0,1]$ is the ratio of observed and intrinsic degree of linear polarization. We present $\xi$ as a function of $\delta$ in Fig.~\ref{fig_depol} for the rest-frame frequencies of 243\,GHz and 354\,GHz, corresponding to rest-frame wavelengths of 1.23\,mm and 0.85\,mm, respectively. Evidently, it is possible to derive $\delta$ if $\xi$ is known for a given rest-frame wavelength. However, this requires information on the intrinsic polarization of 3C\,84 that is not available. Instead, we have to restrict ourselves to the statement that we observe ``substantial'' depolarization at wavelengths as short as 0.85\,mm. When defining substantial depolarization as $\xi\leq0.5$, we are able to derive $\delta\gtrsim4\times10^5$\,rad\,m$^{-2}$ and $\delta\gtrsim8\times10^5$\,rad\,m$^{-2}$ at 243\,GHz and 354\,GHz, respectively. In agreement with the crude estimate from Eq.~\ref{eq_depth}, we see indication for $\Delta{\rm RM}\approx10^6$\,rad\,m$^{-2}$ and, accordingly, $|{\rm RM}|\gtrsim10^6$\,rad\,m$^{-2}$.

Our estimate of RM makes it possible to probe the physical conditions in the parsec-scale environment of 3C\,84 by means of Eq.~\ref{eq_rm}. For $|{\rm RM}|\gtrsim10^6$\,rad\,m$^{-2}$, we find $\langle B_{||} n_e l \rangle\gtrsim1$, with $\langle...\rangle$ indicating the line-of-sight average of the enclosed expression. As pointed out by Taylor et al. \cite{taylor2006}, the effective line-of-sight extension of the Faraday screen can be estimated to $l\approx10$\,pc. As for assessing the electron density, we have to address the various components that contribute. First, the particle density on spatial scales $\lesssim$4\,kpc has been found to be $n_e\approx270$\,cm$^{-3}$ based on optical spectroscopy of [\ion{S}{ii}] lines (Heckman et al. \cite{heckman1989}). Second, for the immediate vicinity ($\lesssim$5\,pc) of the core and the northern counter-jet densities $n_e\gtrsim2000$\,cm$^{-3}$ have been derived from radio continuum spectroscopy (O'Dea, Dent \& Balonek \cite{odea1984}; Walker \& Anantharamaiah \cite{walker2003}). Overall, we may assume an average particle density on the order of $n_e\approx1000$\,cm$^{-3}$, in agreement with particle densities commonly found in the narrow line regions of Seyfert galaxies (e.g., Koski \cite{koski1978}; Bennert et al. \cite{bennert2006}). From the combined information on $l$ and $n_e$ we conclude on the presence of magnetic fields with $B_{||}\gtrsim100\,\mu$G.

Even though the values we derive for RM and $B_{||}$ are fairly high, they are by no means extraordinary. Values of $|{\rm RM}|\approx5\times10^5$\,rad\,m$^{-2}$ can be seen also in the radio nucleus of the Seyfert~1 galaxy 1637+574 (Trippe et al. \cite{trippe2012}) and, most notably, in the Galactic centre radio source Sagittarius~A* (Marrone et al. \cite{marrone2006}; Macquart et al. \cite{macquart2006}). Magnetic fields with $B_{||}\approx100\,\mu$G are compatible to those found in the centre of the Milky Way, where field strengths in the range $\approx$100\,$\mu$G -- 1\,mG have been reported (e.g., Morris \& Yusef-Zadeh \cite{morris1989}; Ferri\`ere \cite{ferriere2009}, and references therein). Accordingly, the parsec-scale environment of 3C\,84 is not unusual with respect to the values of its parameters $n_e$ and $B_{||}$ but -- if at all -- with respect to its spatial variability as seen in projection. The total extension of the radio source is $\lesssim$10\,pc, and observations have resolved structure on scales less than one parsec (e.g., Krichbaum et al. \cite{krichbaum1992}; Taylor et al. \cite{taylor2006}). Accordingly, RM fluctuations with amplitudes on the order of $\Delta{\rm RM}\approx10^6$\,rad\,m$^{-2}$ have to occur -- in projection -- on spatial scales $\lesssim$1\,pc in order to warrant efficient depolarization. This may partially be caused by the fact that observations are affected by contributions from at least two components of interstellar matter with potentially different matter distributions and magnetic field geometries: the foreground ionized gas and the gas associated with the accretion disk. This would also be consistent with the observations by Taylor et al. \cite{taylor2006}; they observed some localized linear polarization in the jet moving toward the observer with an apparent expansion velocity of $\approx0.5c$ (e.g., Asada et al. \cite{asada2006}), i.e. from a source component that would be less affected by an accretion disk and that could have penetrated the Faraday screen partially.

A priori, we may expect some loss of linear polarization by Faraday conversion from linear to circular polarization (e.g., Jones \cite{jones1988}). From single-dish observations, degrees of circular polarization from $m_C\approx0.2$\% at 4.8\,GHz (Aller et al. \cite{aller2003}) up to $m_C\approx0.5$\% at 86\,GHz (Agudo et al. \cite{agudo2010}) have been reported. VLBA imaging at angular resolutions of $\approx$0.8\,mas unveils localized circular polarization in the core with $m_C\approx3$\% at 15\,GHz (Homan \& Wardle \cite{homan2004}). In addition, Homan \& Wardle \cite{homan2004} find for the core polarization $m_C\propto\nu^{-0.9}$; accordingly, polarization levels of $m_C\lesssim0.3$\% may be expected at frequencies $\geq$239\,GHz. From the combined circular polarization information available, we conclude that Faraday conversion contributes only marginally to the loss of linear polarization.

\section{Summary and Conclusions}

We report the results of a search for linear polarization in the active nucleus of the Seyfert~2 galaxy 3C\,84. Observations were carried out with the IRAM Plateau de Bure Interferometer at observatory-frame frequencies of 239\,GHz and 348\,GHz, corresponding to rest-frame frequencies of 243\,GHz and 354\,GHz. We applied Earth rotation polarimetry as the principal tool of analysis. Our study arrives at the following conclusions:

\begin{enumerate}

\item  We do not detect linear polarization. Our analysis finds $3\sigma$ upper limits on the degree of polarization of 0.5\% and 1.9\% at 239\,GHz and 348\,GHz, respectively.

\item  Faraday conversion from linear to circular polarization can be expected on levels of $\lesssim$0.3\%. Accordingly, this mechanism provides only a marginal contribution to the loss of linear polarization observed.

\item  Assuming depolarization by a local Faraday screen, we constrain the rotation measure, as well as the fluctuations therein, to be $\gtrsim10^6$\,rad\,m$^{-2}$. From this we estimate line-of-sight magnetic field strengths of $\gtrsim100\mu$G. These values are fairly high but consistent with observations of other galactic nuclei, most notably the centre of the Milky Way.

\item  Given the physical dimensions of 3C\,84 and its observed structure, the Faraday screen appears to show prominent small-scale structure, with $\Delta{\rm RM}\gtrsim10^6$\,rad\,m$^{-2}$ on projected spatial scales $\lesssim$1\,pc.

\end{enumerate}

Our study underlines the power of polarization observations for studies of the parsec-scale environment of 3C\,84 in special and AGN in general. In order to further probe the Faraday screen of 3C\,84, it will be necessary to obtain observations that actually detect linear polarization on levels $>$1\%. Accordingly, observations at ever higher frequencies, well into the sub-mm/radio regime, should be done. In addition, high-resolution radio interferometric mapping, e.g. with the Global Millimetre-VLBI Array (GMVA), may resolve localized polarized structure and provide new insights into local physical conditions.

\section*{Acknowledgments}

We are grateful to the entire PdBI team for carrying out the observations and technical support. The data analysis made use of the GILDAS software package (\url{http://www.iram.fr/IRAMFR/GILDAS/}) developed and maintained by the GILDAS team. We also applied the software package DPUSER (\url{http://www.mpe.mpg.de/~ott/dpuser/index.html}) developed and maintained by Thomas Ott at MPE Garching.

\bsp

\label{lastpage}

\end{document}